\documentclass[letterpaper,twocolumn,10pt]{article}
\usepackage{usenix2021}
\usepackage[utf8]{inputenc}
\usepackage{xcolor}
\usepackage{xspace}
\usepackage{booktabs}

\newif\ifremoveall{}

\removeallfalse{}

\ifremoveall{}

\newcommand{\moritz}[1]{\textbf{\emph{ #1 \colorbox{yellow}{[Moritz]}}}}
\newcommand{\srdjan}[1]{\textbf{\emph{ #1 \colorbox{blue}{\textcolor{white}{[Srdjan]}}}}}
\newcommand{\shweta}[1]{\textbf{\emph{ #1 \colorbox{green}{[Shweta]}}}}
\newcommand{\ivan}[1]{\textbf{\emph{ #1 \colorbox{magenta}{[Ivan]}}}}
\newcommand{\friederike}[1]{\textbf{\emph{ #1 \colorbox{cyan}{[Friederike]}}}}

\newcommand{\todo}[1]{}
\renewcommand{\todo}[1]{{\color{red} TODO: {#1}}}

\else
\newcommand{\moritz}[1]{}
\newcommand{\srdjan}[1]{}
\newcommand{\friederike}[1]{}
\newcommand{\shweta}[1]{}
\newcommand{\ivan}[1]{}
\newcommand{\todo}[1]{}
\newcommand{\todoref}[1]{}
\newcommand{\citneed}[1]{}
\fi

\newcommand{\sapp}{{sapp}\xspace}
\newcommand{\Sapp}{{Sapp}\xspace}
\newcommand{\sapps}{{sapps}\xspace}
\newcommand{\LOS}{{LOS}\xspace}
\newcommand{\lapp}{{legacy app}\xspace}

 \usepackage{graphicx}
\usepackage{url}

\usepackage{caption}
\usepackage{subcaption}

\usepackage{amssymb}
\usepackage{pifont}
\newcommand{\cmark}{\ding{51}}
\newcommand{\xmark}{\ding{55}}

\title{Sovereign Smartphone:\\ To Enjoy Freedom We Have to Control Our Phones}
\author{
{\rm Friederike Groschupp \quad Moritz Schneider \quad Ivan Puddu \quad Shweta Shinde \quad Srdjan Capkun}\\
{ETH Zurich}
} 
\begin{document}

\maketitle

\begin{abstract} 
The majority of smartphones either run iOS or Android operating
systems. This has created two distinct ecosystems largely controlled
by Apple and Google---they dictate which applications can run, how
they run, and what kind of phone resources they can access. Barring
some exceptions in Android where different phone
manufacturers may have influence, users, developers, and governments
are left with little to no choice. Specifically, users need to entrust
their security and privacy to OS vendors and accept the functionality constraints they impose. Given the wide use of Android and
iOS, immediately leaving these ecosystems is not practical,
except in niche application areas. In this work, we draw attention to 
the magnitude of this problem and why it is an undesirable
situation. As an alternative, we advocate the development of a new
smartphone architecture that securely transfers the control back to
the users while maintaining compatibility with the rich existing
smartphone ecosystems. We propose and analyze one such design based on
 advances in trusted execution environments for ARM and
RISC-V.
\end{abstract}
 \section{Introduction}
Smartphones are the centerpiece of most people's digital life.
However, they do not offer the same flexibility as PCs, where users
can install and run arbitrary software.
Smartphone manufacturers and vendors of major operating systems such as
iOS and Android\footnote{In the following, Android
refers to the Google-controlled variant. While there is the Android Open Source Project (AOSP), it does not come with core features such as Google Mobile Services, which are closed-source and proprietary to Google~\cite{arstechnica_2018,google_gms}} restrict which apps can be run on smartphones, type of
peripheral access, and data access. While Android allows side-loading,
users are still limited in the way they can run apps and the kind of
access they have. There are several examples where, in order to
protect the users from developers, Apple and Google limit apps' access
to peripherals and data even if the users would allow such
access~\cite{reuters_2020,androidBackground}.

Smartphone manufacturers and OS vendors can use such control to
optimize performance, provide a good user experience, and protect
users from malicious apps.
At the same time, these companies become arbiters and gatekeepers. Recent examples show that this is not a minor issue. In the
case of contact tracing apps, Apple (and to some extent Google)
restricted access that government apps can have to Bluetooth beacons,
citing privacy and performance concerns. This restricted the design space and performance of contact tracing apps in several
countries~\cite{theguardian_2020_a,theguardian_2020_b,reuters_2020}.
After recent developments in the US, the Parler app has been removed from
Apple and Google app stores~\cite{forbes_2021} and Google banned the app of a traditional Danish children's program after the company deemed
its content unsuitable~\cite{reuters_2020_b}. Apple and Google
policies required in-app purchases to use in-store payments, resulting in these companies being accused of gatekeeping in several jurisdictions~\cite{thenewyorktimes_2020}. Users,
developers, and governments have therefore faced restrictions under
the current model. Most notably, users are unable to freely use their
smartphones---they are subject to several limitations that they do
not face on their PCs. 

This is clearly an undesirable situation. Even if the current OSs
offer some leeway, like side-loading, they can and have taken it away
from users and developers~\cite{theverge_2021}. Few companies should not be in a
position to have such control. As much as possible, control over the
devices should be handed back to the users who own the phones and
whose data the phones primarily hold. 

This is a complex technical, legal, and societal issue that cannot be
resolved by technical means alone~\cite{digital-markets-act-eu,
digital-markets-act-21}. However, new smartphone designs can be a part
of the solution to this problem. Ideally, a \emph{sovereign
smartphone} will give the user full control of the software, hardware
(e.g., peripherals), and data on the phone. One obvious solution is to
replicate the PC model for smartphones. Projects that allow this
already exist~\cite{pinephone,libremphone}. However, they require the
users to switch to other app stores. More importantly, they cannot
easily replicate the functionality and protections that existing OSs
offer. 
Virtualization-based
solutions~\cite{andrus2011cells,barham2003xen} would hand the control
to the hypervisors. However, such privileged software can inspect and modify
the OS and app memory, therefore, removing control from the users and
preventing OS vendor from protecting their ecosystems. Furthermore, removing control from the OS vendor only to hand it to the smartphone manufacturers, which would typically control such hypervisors, would not solve the underlying issues.
 
We propose a new design that demonstrates the feasibility of
building a \emph{sovereign smartphone}. We give control back to the
users, without taking away the functionality and security from the
existing operating system vendors and smartphone manufacturers. Such
design allows companies to offer comprehensive protection to the users
and software on the platform without restricting the users' ability
to run software and access platform resources. 
The sovereign smartphone offers flexibility in the deployment of new apps
and functionalities and aligns the interests of
users, phone manufacturers, cellular network operators, and even OS
vendors.

Our design builds on the recent advances in trusted execution
environments (TEEs). We discuss how it can be implemented on top of ARM and
RISC-V TEE architectures. We outline several critical technical
challenges in realizing sovereign smartphones. \section{Need for Sovereignty} 
\label{sec:problem_statement}

\subsection{Prominent Examples of Restrictions} 
\hspace{\parindent}\textbf{Restricted resource access.}
Bluetooth-based contact tracing apps periodically receive and process
Bluetooth Low Energy (BLE) beacons to measure the distance to other
smartphones and register contact. Due to concerns about privacy and power consumption, iOS and Android did not permit apps running in the background to freely broadcast and receive BLE beacons, effectively disallowing BLE-based approaches preferred by some countries~\cite{theguardian_2020_a, theguardian_2020_b, reuters_2020}. Instead, the introduction of such contract tracing apps almost entirely depended on Apple and Google implementing and providing an API for a particular decentralized contact tracing approach~\cite{reuters_2020,apple_2020}. Clearly, these companies can always disable this API.

\textbf{Censored app availability.}
App store providers control which apps are offered on their app stores. They may
reject or remove apps based on company policies~\cite{google_guidelines, apple_guidelines}, public pressure, or executive
authority. There have been several notable instances of apps being
 removed from official Apple and Google app stores. Examples include the ad-blocker Adblock Fast \cite{theverge_2016}, the game Fortnite over a feud on payment restrictions~\cite{theverge_2020}, the right-wing
app Parler after civil unrests in the US~\cite{cnn_2021, forbes_2021}, and the
HKmap Live app used by protesters during the Hong Kong 
protests~\cite{bbc_2019}. If an app is not available through the official
app stores, iOS users cannot easily install it.\footnote{On iOS, custom apps can be installed with a special developer account and the app's source code through XCode.} On Android, developers may
circumvent this restriction by offering their app as a side-loaded
package. 
Even if technically users can side-load apps, iOS and Android can easily take this privilege away, either by selectively blocking apps or preventing any side-loading~\cite{theverge_2021}. Furthermore, it might be technically feasible that OS vendors are legally compelled to disclose if banned apps are running on a user's phone~\cite{apple-gov-info-request}.

\textbf{Forced Ecosystem.} \todo{title:Gatekeeping?}
When an app is distributed through the official app stores, all
payment transactions, even in-app purchases, must be processed through
the respective billing service offered by Apple and
Google~\cite{thenewyorktimes_2020}. This forces developers to use a specific payment API and subdues them to any fees imposed, as they cannot easily avoid offering their apps on these app stores~\cite{androidfee,cnbc_2020}.

\textbf{Data privacy concerns.}
Users face uncertainty about when and what kind of data is collected
by their phone, for example through the phone sensors, and how it is
processed~\cite{businessinsider_2019}. While the OS allows the user to manage peripheral
permissions for apps or to disable access to some resources globally, e.g., by turning off GPS
or Bluetooth, they have to trust the OS.
Intentional or unintentional misuse of such OS-level privileges or opaque policies can
put the user data in danger without the user's
knowledge. For example, Google gathered location information even when the location history feature was turned off~\cite{apnews_2018, schmidt2018google}.

\subsection{Advantages of Existing Ecosystems}

OS vendors invest large amounts of effort, money, and thought in
developing a phone OS. By acting as a central authority in their
ecosystems, they simplify management and engineering tasks. For instance, 
they can quickly react to zero-day vulnerabilities or new classes of
attacks~\cite{lipp2018meltdown,androidmeltdown}. They can avoid fragmentation of
their ecosystem because they provide a unified system, consistent
APIs, and central app stores. 

Apple and Google aim to protect their users' security and privacy against third parties. Specifically, they vet apps and weed out potential malware before releasing them on the official store. To enhance security, the OSs restrict user permissions, enforce strict inter-app isolation, and suppress direct access to certain peripherals. Further, they provide useful security services, e.g., protection for digital copyright content, Google SafetyNet, secure OS boot, OS tamper detection, and app-specific data protection. In summary, a tightly controlled system preserves users' and apps' usability, security, and privacy while providing a good developer experience.

\subsection{Bypassing App Stores and OS Vendors}
\hspace{\parindent}\textbf{Rooting/Jailbreaking.} 
Users can gain root permissions on their phones by bypassing security mechanisms or exploiting vulnerabilities. They can run
privileged apps and circumvent hardware manufacturer or OS vendor restrictions. However, this approach is not user-friendly, might brick the phone, and voids device warranty~\cite{sun2015android, applesupport_2018}.

\textbf{Unlocking the bootloader.} 
Users can unlock the bootloader on some Android phones and install a different OS. In theory, the user then has full control over the OS functionality, drivers, and the apps they want to install. However, alternative OSs lack compatibility with common apps and key proprietary functionality~\cite{arstechnica_2018}. In addition, even though unlocking the bootloader is supported by some hardware vendors, it still voids the warranty in many cases and irreversibly prevents particular apps from running, such as Samsung's security framework Knox~\cite{samsung_knox}.

\textbf{Side-loading.} 
Android allows side-loading, i.e., the installation of code distributed by other means than the official app store~\cite{samsung_2019sideloading}.
However, the user depends on OS vendor support for this feature, which can be potentially discontinued~\cite{theverge_2021}. Furthermore, this approach does not allow unrestricted access to resources. For example, side-loading a contact tracing app would not have resolved the lack of functionality at the OS level. 

 \section{Sovereign Smartphones}
\label{sec:sovereign_smartphones}

Reducing OS vendors' control in the phone market necessitates shifting
it to other entities, i.e., users, governments, or independent
oversight committees. We argue that the users, as device owners and as
those whose information is primarily processed on the devices, should
be in control of their data and able to freely use their devices
without OS vendor restrictions.

Giving users full control over their phone has some trade-offs. First,
it would simplify illegal activities on phones which could not be
prevented by authorities. However, criminals already have ways of
circumventing these limitations, such as using PCs or open stack
devices. Law enforcement has other ways of control, e.g., through
ISPs. Second, an average phone user may not have the technical
expertise or necessity to configure a sovereign phone. Such a user can
use configurations offered by a trusted party (e.g., independent
committees). Alternatively, they can continue using an existing OS as
is, if the sovereign phone platform is compatible with current OSs.
While this essentially delegates the control from the user to another
entity, the choice is still up to the user.

\subsection{Desired Properties}
\label{sec:properties}

We envision that a sovereign smartphone should maintain compatibility
with current OSs such as iOS and Android, referred to as
\textit{legacy operating system (\LOS)} in the following. Users can
continue to execute existing apps, which are hereafter referred to as
\textit{\lapp{}s}. We introduce the notion of \textit{sovereign app}
(\sapp), an app that runs outside the \LOS. A user can run a \sapp
when (i) it is not available in the \LOS app store and side-loading
might pose a security risk to the \LOS; (ii) they do not trust the
\LOS; or (iii) the \sapp needs access to resources prohibited by the
\LOS.

\textbf{(P1) Full user control.} 
The user can assign certain shares of compute time to \LOS and \sapps.
The user can deny or grant exclusive peripheral access to  \LOS or
\sapps.

\textbf{(P2) \LOS protection.}
The \LOS and \lapp{}s continue to have the same guarantees as they
would in the absence of \sapps. The \LOS is still in charge of
protecting \lapp{}s and enforcing permissions within its domain,
\lapp{}s cannot directly access resources or bypass \LOS protections,
and the \LOS can inspect and censor \lapp{}s. The \LOS and \lapp{}s
are guaranteed to maintain their existing confidentiality, integrity,
and availability even in the presence of \sapps. 

\textbf{(P3) \Sapp protection.}
Each \sapp has confidentiality, integrity, and availability guarantees
in the presence of \LOS, \lapp{}s, and other \sapps.

\textbf{(P4) Execution without leaking \sapp identity.} 
Neither the \LOS nor \sapps should have access to information that
reveals the identity of other \sapps on the device.

\textbf{(P5) Limited trusted computing base (TCB).}
The design should assume only a small amount of code to be trusted and
bug-free. The TCB should be around few thousand LoC i.e., within the
realm of formal verification.

Satisfying P1--P5 enables several use-cases. For example, a contact
tracing \sapp can be granted exclusive access to the Bluetooth card
and guaranteed a certain periodic runtime by a user policy (P1). As
another example, a user could install a messaging app that is not
offered on the official app stores. They are assured that \sapp
confidentiality and integrity is protected (P3) on the phone.
Technically, the legislation cannot force the LOS, other \lapp{}s or
\sapps on the phone to disclose that the user has installed or is
running a particular \sapp (P4). In both instances, the user can
continue to use the \LOS with the same confidence in its properties as
before (P2).

\subsection{Strawman Approaches}
\label{sec:strawman}

\begin{table}
\small 
\centering
\label{tab:naive_solutions}
\begin{tabular}{lccccc} \toprule
\bf{Solution}     & \bf{P1} & \bf{P2} & \bf{P3} & \bf{P4} & \bf{P5} \\ \midrule
Root permissions  & \xmark  & \xmark  & \xmark  & \xmark  & \xmark  \\ 
Extending LOS     & \cmark  & \xmark  & \xmark  & \xmark  & \xmark  \\ 
Hypervisor        & \xmark  & \cmark  & \cmark  & \cmark  & \xmark  \\ 
Traditional TEEs  & \xmark  & \cmark  & (\cmark)  & \xmark  & \cmark  \\ \midrule
\bf{Our Design}   & \cmark  & \cmark  & \cmark  & \cmark  & \cmark  \\ \bottomrule
\end{tabular}
\caption{Properties provided by different solutions.
P3 (\cmark) implies  confidentiality and integrity, without availability. 
}
\vspace{-1em}
\end{table}

\hspace{\parindent}\textbf{Giving users full root permission} to
install an \sapp as an app in the LOS can be a legitimate alternative
to rooting. However, as the LOS is still mostly in charge, this
approach satisfies neither of our properties: The user cannot assign
resources (P1), the LOS and \sapps are not protected from each other
(P2, P3), and the LOS can monitor which \sapps are installed and
running (P4). 

\textbf{Extending the LOS} with the capability of loading
user-provided kernel modules, i.e., drivers, would allow  users to
configure resource access (P1). However, the LOS and \sapps would not
be isolated from each other. Furthermore, the LOS would still be in
charge of launching and scheduling all apps, allowing it to inspect
\sapps and deny service to \sapps (P3, P4). We cannot trust the entire
\LOS (P5).

\textbf{Hypervisors} can virtualize resources and isolate the LOS in a
VM~\cite{barham2003xen,andrus2011cells}, thus fulfilling P2 and P3.
\Sapp{}s can then execute in their own VM. However, the hypervisor
itself executes with the highest privileges. It can directly inspect
VM memory and interfere with its execution. Moreover, typical
hypervisors have large code bases, up to 1M LOC
(P5)~\cite{barham2003xen}. More importantly, hypervisor vendors may
resort to similar tactics as current OS vendors and curtail user's
control over their device. 

\textbf{Executing \sapps in traditional TEE enclaves} can provide
confidentiality and integrity to \LOS and \sapp{}s
(P2)~\cite{lee2020keystone,costan2016intel} with a minimal TCB
(P5)~\cite{lebedev2019sanctorum}. As the \LOS controls scheduling,
availability is only guaranteed for the \LOS, not for \sapp{}s (partial P3). Furthermore, the \LOS can fingerprint a \sapp or detect
which \sapps are executed (P4). Current TEE proposals do not allow
access to system resource without mediation by the LOS (P1).
 \section{Our Design}

\begin{figure}
    \centering
    \begin{subfigure}[b]{0.45\linewidth}
        \centering
        \includegraphics[width=0.95\linewidth]{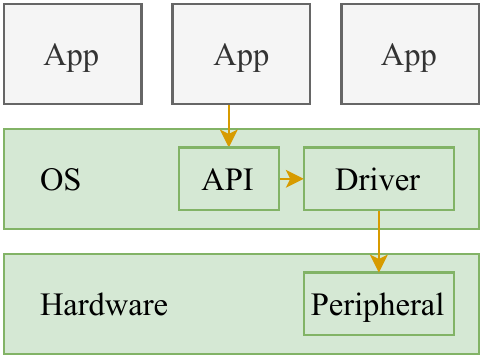}
        \caption{Current smartphones}
    \end{subfigure}
    \,
    \begin{subfigure}[b]{0.45\linewidth}
        \centering
        \includegraphics[width=0.95\linewidth]{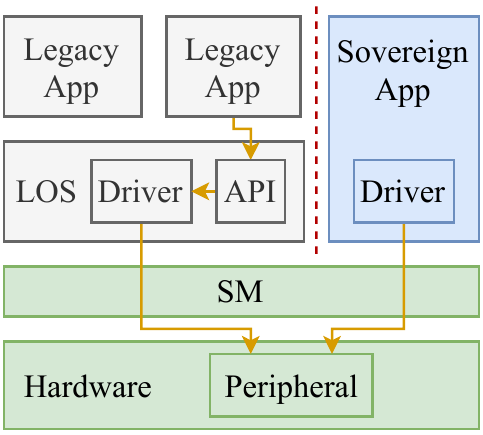}
        \caption{Our proposal}
    \end{subfigure}
    \caption{Software architecture of (a) current smartphones and (b) our proposal. Green denotes trusted components, grey and blue are  mutually-untrusted isolation domains.}
    \label{fig:design}
    \vspace{-1em}
\end{figure}

Our approach is based on a TEE design with a small security monitor (SM) containing the software TCB~\cite{lee2020keystone} (see \autoref{fig:design}). The SM runs with special privileges and is responsible for configuring isolation and peripheral access during context switches between the LOS and \sapp{}s. 
However, there are key differences to TEEs. First, the SM is not omnipotent in our approach: It can manage \sapp{}s and the LOS, but not inspect their memory or interfere with their execution. Second, \sapp{}s are able to access system resources such as Bluetooth without the LOS being able to inspect or interfere. And third, we aim to provide  availability guarantees to both the LOS and \sapp{}s. 

\textbf{Management without inspection.}
Maintaining existing security guarantees and functionality of the LOS (P2) is critical. Thus, we try to reduce the SM's capabilities by removing its access permissions to the LOS or \sapp{}s. Similar solutions exist in cloud computing, where a hypervisor cannot inspect the private memory of its guests~\cite{amdsev,inteltdx}. 
Traditional ARM or RISC-V platforms have partial support for 
such hardware-based protection.
We discuss how to enable this primitive for smartphones in 
Section~\ref{sec:tech_feasibility}. 

\textbf{Resource sharing.}
Traditionally, OSes manage system resources. Adding these management capabilities to the SM would bloat the TCB and only shift problems to a lower layer. Instead, we propose the \LOS to remain in charge of management, but it can concede control of a peripheral to a \sapp. From that point on, the SM enforces the \sapp's exclusive control over the peripheral by configuring a hardware access control mechanism to restrict access to the peripheral memory ranges. In a mobile phone, all peripherals are set statically at design time. Their memory-mapped addresses and version numbers are included in a device tree file burned into ROM on the SoC. Thus, the SM can consult the trusted device tree to ensure that a particular peripheral is exclusively controlled by a \sapp{}. Note that this approach only allows one \sapp{} or the LOS to  exclusively access a single peripheral at a time. We are investigating proposals that allow for more flexibility.

\textbf{Mutual Progress.}
Ensuring availability can be achieved by adding a scheduler to the SM.
However, such scheduling functionality must be immutable, i.e., 
an attacker cannot update it without getting detected. 
On the other hand, the LOS has a sophisticated scheduler that is well-suited for legacy apps and hardware but is not trusted.
Therefore, we propose a middle-ground: The user or the \sapp{} can specify policies, e.g., a \sapp{} should be assigned a certain amount of runtime during a specified interval. The LOS performs the scheduling. The SM verifies that it is done as per the \sapp{}- or user-defined policies by recording the runtime that was granted to the \sapp{} upon context switches. Since policy verification may be expensive for complex policies, we propose periodic checks. The SM merely verifies that these periodic checks have been performed regularly on every context switch. If the SM detects violations to the policies or no periodic checks, it can impose sanctions (e.g., user notification or  locking the device operation). We assume that this is a lose-lose situation that all sides want to prevent. Moreover, once informed, the user can replace the  uncooperative LOS.

\textbf{Attestation.}
In our proposal, a local user wants to attest and verify the code running within a \sapp{}. In addition, the user wants to verify the scheduling policies and resource accesses permissions of the \sapp{} are initialized correctly. 
Remote attestation of traditional TEEs 
has the basic mechanisms to provide such guarantees. We will augment it to include \sapp specific information. Moreover, our SM can directly interact with the user through a secure user interface to signal successful attestation. 

\subsection{Feasibility on RISC-V and ARM} \label{sec:tech_feasibility}

On RISC-V, the SM runs in machine mode. It configures physical memory protection (PMP) entries to isolate \sapp{}s and the \LOS running in the less-privileged user and supervisor modes respectively~\cite{lee2020keystone}. 
We will use existing hardware mechanisms to restrict the SM from inspecting \sapp{} and \LOS memory. Specifically, the SM locks the PMP entries after \sapp{} creation with a sticky bit (which is only cleared on full system reset) such that it does not have access to the respective memory regions~\cite{riscv2019privspec}. 

On ARM, the SM runs in the secure world. It 
configures the TrustZone address space controller (TZASC)~\cite{brasser2019sanctuary} to 
provision isolated execution environments in the normal world for the LOS and \sapp{}s. To restrict the SMs capability to inspect memory, we are investigating a combination of features from two different TZASCs: TZC-380 for locking configurations~\cite{arm380} and TZC-400 for per-core memory isolation~\cite{arm400}. Peripheral access can be enforced by leveraging the TZASC as proposed in~\cite{lentz2018secloak}.

 \subsection{Analysis}

\hspace{\parindent}\textbf{Threat Model.} 
In general, we assume the SM and the hardware to be trusted. A malicious SM could deny running certain \sapps or monitor the user's activities. Therefore, the SM should remain as small as possible and should potentially be implemented purely in hardware. 

One has to consider at least two perspectives when discussing the threat model of the sovereign smartphone. First, the LOS and \sapp{}s consider a malicious user against whom they want to protect their internal secrets. This means that a local physical adversary is in scope in this scenario. However, we note that current smartphones also consider such an adversary. 
Second, from the user's perspective, the LOS is considered malicious, e.g., the LOS might be under societal or even legal pressure to censor some \sapp{}s, or even report the user to law enforcement for the usage of an illegal \sapp{}. In the worst case, the LOS is entirely malicious and tries to leak the users' data. In this scenario, the user remains in physical possession of the phone, and thus, there is no physical adversary. We also assume that the foremost goal of the LOS is that users keep running it, as a complete refusal to cooperate might hurt both the user and the LOS.

\textbf{Security Analysis.} \label{sec:security_analysis}
The user needs to be able to assign system resources to \sapp{}s (P1). In our approach, the LOS can concede control of individual peripherals to \sapp{}s, which the user can then verify. The SM enforces exclusive access to these peripherals as per the trusted device tree burned into ROM. 

As in traditional TEEs, the \sapp{}s or the LOS cannot access each other's private memory (P2 and P3). However, the SM can always read all memory on the system. To address this gap, we plan to remove the SM's capability to inspect memory by using once-set-never-unset mechanisms in  the hardware (until hard-reset)~\cite{riscv2019privspec}. 

While our proposal does not grant absolute availability, it guarantees mutual progress of the LOS and \sapp{}s. Moreover, we prevent selective denial-of-service of an individual \sapp{}. The SM will notice such violations by the OS and take necessary actions to disincentivize this.

We require that \sapps can be installed and executed on our platform without being identified by entities other than the user and the SM (P4). We stress that this is a known hard problem~\cite{meyerson2004complexity} and may be impossible for \sapp{}s with clearly distinct resource usage patterns. However, there are multiple probabilistic approaches leveraging obfuscation to hide \sapp{} identities. We also note that the anonymity property might conflict with the flexibility, policy expressiveness, and the attestation mechanism. 

Many \sapp{}s and the SM will require user interaction. The security of this user interaction is critical to prevent various user-interface attacks~\cite{fratantonio2017cloak,chen2014peeking}. There have been various studies on the effectiveness of security indicators~\cite{schechter2007emperor, whalen2005gathering}, and recent work has proposed secure LEDs~\cite{eskandarian2019fidelius} or extra buttons~\cite{lentz2018secloak}. However, it is unclear if these proposals are fully applicable to our scenario, or if there are more lingering issues. 

Thus, designing a sovereign smartphone that fulfills our outlined properties presents a unique set of non-trivial technical challenges. Our analysis shows that, although not straightforward, such a design is feasible. \section{Outlook}

In our proposal, we lay the foundation for a sovereign smartphone architecture. It combines handing users control over their devices with advantages of current ecosystems in a secure manner. It highlights important challenges that need attention from the technical community.

\bibliographystyle{plain}

\begin{thebibliography}{10}

\bibitem{amdsev}
AMD.
\newblock {AMD} memory encryption.
\newblock Technical report, AMD, 2016.

\bibitem{andrus2011cells}
Jeremy Andrus, Christoffer Dall, Alexander~Van't Hof, Oren Laadan, and Jason
  Nieh.
\newblock Cells: A virtual mobile smartphone architecture.
\newblock In {\em Proceedings of the Twenty-Third ACM Symposium on Operating
  Systems Principles}, pages 173--187, 2011.

\bibitem{apnews_2018}
{AP News}.
\newblock Google tracks your movements, like it or not, Aug 2018.
\newblock \url{https://apnews.com/article/828aefab64d4411bac257a07c1af0ecb}.

\bibitem{apple_guidelines}
Apple.
\newblock App store review guidelines.
\newblock \url{https://developer.apple.com/app-store/review/guidelines}.

\bibitem{apple_2020}
Apple.
\newblock Apple and {G}oogle partner on {COVID}-19 contact tracing technology,
  Apr 2020.
\newblock
  \url{https://www.apple.com/newsroom/2020/04/apple-and-google-partner-on-covid-19-contact-tracing-technology}.

\bibitem{apple-gov-info-request}
Apple.
\newblock Privacy - government information requests - apple.
\newblock \url{https://www.apple.com/privacy/government-information-requests/},
  2021.

\bibitem{applesupport_2018}
{Apple Support}.
\newblock Unauthorized modification of i{OS} can cause security
  vulnerabilities, instability, shortened battery life, and other issues, Jun
  2018.
\newblock \url{http://support.apple.com/kb/HT3743}.

\bibitem{arm380}
ARM.
\newblock {CoreLink TrustZone} address space controller {TZC}-380 technical
  reference manual.
\newblock Technical report, ARM, 2010.

\bibitem{arm400}
ARM.
\newblock {ARM CoreLink TZC-400 TrustZone} address space controller technical
  reference manual.
\newblock Technical report, ARM, 2014.

\bibitem{arstechnica_2018}
{Ars Technica}.
\newblock Googles iron grip on {A}ndroid: Controlling open source by any means
  necessary, Jul 2018.
\newblock
  \url{https://arstechnica.com/gadgets/2018/07/googles-iron-grip-on-android-controlling-open-source-by-any-means-necessary/}.

\bibitem{barham2003xen}
Paul Barham, Boris Dragovic, Keir Fraser, Steven Hand, Tim Harris, Alex Ho,
  Rolf Neugebauer, Ian Pratt, and Andrew Warfield.
\newblock Xen and the art of virtualization.
\newblock {\em ACM SIGOPS operating systems review}, 37(5):164--177, 2003.

\bibitem{bbc_2019}
{BBC News}.
\newblock Apple drops {H}ong {K}ong police-tracking app used by protesters, Oct
  2019.
\newblock \url{https://www.bbc.com/news/business-49995688}.

\bibitem{brasser2019sanctuary}
Ferdinand Brasser, David Gens, Patrick Jauernig, Ahmad-Reza Sadeghi, and
  Emmanuel Stapf.
\newblock Sanctuary: {ARM}ing {T}rust{Z}one with user-space enclaves.
\newblock In {\em NDSS}, 2019.

\bibitem{businessinsider_2019}
{Business Insider}.
\newblock A security expert found that {A}pple's latest i{P}hone can still
  track your location data, even if you toggle it off for every app, Dec 2019.
\newblock
  \url{https://www.businessinsider.com/apple-iphone-11-pro-collects-location-data-krebs-report-2019-12?r=US&IR=T}.

\bibitem{chen2014peeking}
Qi~Alfred Chen, Zhiyun Qian, and Z~Morley Mao.
\newblock Peeking into your app without actually seeing it: {UI} state
  inference and novel {A}ndroid attacks.
\newblock In {\em 23rd USENIX Security Symposium (USENIX Security 14)}, pages
  1037--1052, 2014.

\bibitem{cnbc_2020}
CNBC.
\newblock Fortnite maker: '{A}pple has locked down and crippled' the app store,
  Jul 2020.
\newblock
  \url{https://www.cnbc.com/2020/07/24/epic-games-ceo-tim-sweeney-apple-crippled-app-store-with-30percent-cut.html}.

\bibitem{cnn_2021}
CNN.
\newblock Parler, right-wing social media app, is removed from the {G}oogle
  {P}lay {S}tore, Jan 2021.
\newblock
  \url{https://edition.cnn.com/2021/01/08/tech/parler-google-play-removed/index.html}.

\bibitem{costan2016intel}
Victor Costan and Srinivas Devadas.
\newblock Intel {SGX} explained.
\newblock {\em IACR Cryptol. ePrint Arch.}, 2016(86):1--118, 2016.

\bibitem{digital-markets-act-21}
Fiona Scott~Morton Cristina~Caffarra.
\newblock The european commission digital markets act: A translation | vox,
  cepr policy portal.
\newblock
  \url{https://voxeu.org/article/european-commission-digital-markets-act-translation},
  Jan 2021.

\bibitem{eskandarian2019fidelius}
Saba Eskandarian, Jonathan Cogan, Sawyer Birnbaum, Peh Chang~Wei Brandon,
  Dillon Franke, Forest Fraser, Gaspar Garcia, Eric Gong, Hung~T Nguyen,
  Taresh~K Sethi, et~al.
\newblock Fidelius: Protecting user secrets from compromised browsers.
\newblock In {\em 2019 IEEE Symposium on Security and Privacy (SP)}, pages
  264--280. IEEE, 2019.

\bibitem{digital-markets-act-eu}
{European Commission}.
\newblock The digital markets act: ensuring fair and open digital markets.
\newblock
  \url{https://ec.europa.eu/info/strategy/priorities-2019-2024/europe-fit-digital-age/digital-markets-act-ensuring-fair-and-open-digital-markets_en},
  Jan 2019.

\bibitem{forbes_2021}
Forbes.
\newblock Parler at risk of going offline after bans from {A}mazon, {A}pple and
  {G}oogle, Jan 2021.
\newblock
  \url{https://www.forbes.com/sites/jemimamcevoy/2021/01/10/parler-at-risk-of-going-offline-after-bans-from-amazon-apple-and-google/}.

\bibitem{fratantonio2017cloak}
Yanick Fratantonio, Chenxiong Qian, Simon~P Chung, and Wenke Lee.
\newblock Cloak and dagger: from two permissions to complete control of the
  {UI} feedback loop.
\newblock In {\em 2017 IEEE Symposium on Security and Privacy (SP)}, pages
  1041--1057. IEEE, 2017.

\bibitem{google_guidelines}
Google.
\newblock Developer program policy.
\newblock
  \url{https://support.google.com/googleplay/android-developer/answer/10355942?hl=en}.

\bibitem{google_gms}
Google.
\newblock Google mobile services.
\newblock \url{https://www.android.com/gms/}.

\bibitem{androidmeltdown}
Google.
\newblock Android security bulletin - january 2018, January 2018.
\newblock \url{https://source.android.com/security/bulletin/2018-01-01}.

\bibitem{androidBackground}
Google.
\newblock Media{R}ecorder overview, January 2021.
\newblock \url{https://developer.android.com/guide/topics/media/mediarecorder}.

\bibitem{androidfee}
Google.
\newblock Service fees, 2021.
\newblock
  \url{https://support.google.com/googleplay/android-developer/answer/112622?hl=en}.

\bibitem{inteltdx}
Intel.
\newblock Intel trust domain extensions.
\newblock Technical report, Intel, 2020.

\bibitem{lebedev2019sanctorum}
Ilia Lebedev, Kyle Hogan, Jules Drean, David Kohlbrenner, Dayeol Lee, Krste
  Asanovi{\'c}, Dawn Song, and Srinivas Devadas.
\newblock Sanctorum: A lightweight security monitor for secure enclaves.
\newblock In {\em 2019 Design, Automation \& Test in Europe Conference \&
  Exhibition (DATE)}, pages 1142--1147. IEEE, 2019.

\bibitem{lee2020keystone}
Dayeol Lee, David Kohlbrenner, Shweta Shinde, Krste Asanovi{\'c}, and Dawn
  Song.
\newblock Keystone: An open framework for architecting trusted execution
  environments.
\newblock In {\em Proceedings of the Fifteenth European Conference on Computer
  Systems}, pages 1--16, 2020.

\bibitem{lentz2018secloak}
Matthew Lentz, Rijurekha Sen, Peter Druschel, and Bobby Bhattacharjee.
\newblock Secloak: {ARM} {T}rust{Z}one-based mobile peripheral control.
\newblock In {\em Proceedings of the 16th Annual International Conference on
  Mobile Systems, Applications, and Services}, pages 1--13, 2018.

\bibitem{lipp2018meltdown}
Moritz Lipp, Michael Schwarz, Daniel Gruss, Thomas Prescher, Werner Haas,
  Anders Fogh, Jann Horn, Stefan Mangard, Paul Kocher, Daniel Genkin, et~al.
\newblock Meltdown: Reading kernel memory from user space.
\newblock In {\em 27th {USENIX} Security Symposium ({USENIX} Security 18)},
  pages 973--990, 2018.

\bibitem{meyerson2004complexity}
Adam Meyerson and Ryan Williams.
\newblock On the complexity of optimal k-anonymity.
\newblock In {\em Proceedings of the twenty-third ACM SIGMOD-SIGACT-SIGART
  symposium on Principles of database systems}, pages 223--228, 2004.

\bibitem{pinephone}
Pine64.
\newblock Pine{P}hone.
\newblock \url{https://www.pine64.org/pinephone/}.

\bibitem{libremphone}
Purism.
\newblock Librem 5.
\newblock \url{https://puri.sm/products/librem-5/}.

\bibitem{reuters_2020_b}
Reuters.
\newblock Denmark angry at {G}oogle censorship of some {D}anish content, seeks
  talks, Aug 2020.
\newblock
  \url{https://www.re}uters.com/article/us-google-censorship-denmark-idUSKCN2561TG.

\bibitem{reuters_2020}
Reuters.
\newblock Germany at odds with {A}pple on smartphone coronavirus contact
  tracing, Apr 2020.
\newblock
  \url{https://www.reuters.com/article/us-health-coronavirus-europe-tech/germany-at-odds-with-apple-on-smartphone-coronavirus-contact-tracing-idUSKCN2251MR}.

\bibitem{samsung_2019sideloading}
{Samsung Business Insights}.
\newblock What are the risks of sideloaded {A}ndroid applications?, Nov 2019.
\newblock
  \url{https://insights.samsung.com/2019/11/26/what-are-the-risks-of-sideloaded-android-applications/}.

\bibitem{samsung_knox}
{Samsung Knox}.
\newblock What is a {K}nox warranty bit and how is it triggered?
\newblock
  \url{https://docs.samsungknox.com/admin/knox-platform-for-enterprise/faqs/faq-115013562087.htm}.

\bibitem{schechter2007emperor}
Stuart~E Schechter, Rachna Dhamija, Andy Ozment, and Ian Fischer.
\newblock The emperor's new security indicators.
\newblock In {\em 2007 IEEE Symposium on Security and Privacy (SP'07)}, pages
  51--65. IEEE, 2007.

\bibitem{schmidt2018google}
Douglas Schmidt.
\newblock Google data collection.
\newblock Technical report, Vanderbilt University, 2018.
\newblock
  \url{http://www.dre.vanderbilt.edu/~schmidt/PDF/google-data-collection.pdf}.

\bibitem{sun2015android}
San-Tsai Sun, Andrea Cuadros, and Konstantin Beznosov.
\newblock Android rooting: Methods, detection, and evasion.
\newblock In {\em Proceedings of the 5th Annual ACM CCS Workshop on Security
  and Privacy in Smartphones and Mobile Devices}, pages 3--14, 2015.

\bibitem{theguardian_2020_a}
{The Guardian}.
\newblock France urges {A}pple and {G}oogle to ease privacy rules on contact
  tracing, Apr 2020.
\newblock
  \url{https://www.theguardian.com/world/2020/apr/21/france-apple-google-privacy-contact-tracing-coronavirus}.

\bibitem{theguardian_2020_b}
{The Guardian}.
\newblock {NHS} in standoff with {A}pple and {G}oogle over coronavirus tracing,
  Apr 2020.
\newblock
  \url{https://www.theguardian.com/technology/2020/apr/16/nhs-in-standoff-with-apple-and-google-over-coronavirus-tracing}.

\bibitem{thenewyorktimes_2020}
{The New York Times}.
\newblock Fortnite creator sues {A}pple and {G}oogle after ban from app stores,
  Aug 2020.
\newblock
  \url{https://www.nytimes.com/2020/08/13/technology/apple-fortnite-ban.html}.

\bibitem{theverge_2016}
{The Verge}.
\newblock Google removes {S}amsung's first {A}ndroid ad blocker from the {P}lay
  {S}tore, Feb 2016.
\newblock
  \url{https://www.theverge.com/2016/2/3/10905672/google-samsung-adblock-fast-android-ad-blocker-removal}.

\bibitem{theverge_2020}
{The Verge}.
\newblock Apple just kicked {F}ortnite off the app store, Aug 2020.
\newblock
  \url{https://www.theverge.com/2020/8/13/21366438/apple-fortnite-ios-app-store-violations-epic-payments}.

\bibitem{theverge_2021}
{The Verge}.
\newblock Apple is blocking {A}pple silicon mac users from sideloading i{P}hone
  apps, Jan 2021.
\newblock
  \url{https://www.theverge.com/2021/1/15/22233754/apple-blocking-m1-iphone-app-sideloading}.

\bibitem{riscv2019privspec}
Andrew Waterman, Yunsup Lee, Rimas Avizienis, David~A Patterson, and Krste
  Asanovi{\'c}.
\newblock The {RISC-V} instruction set manual volume ii: Privileged
  architecture.
\newblock {\em EECS Department, University of California, Berkeley}, 2019.

\bibitem{whalen2005gathering}
Tara Whalen and Kori~M Inkpen.
\newblock Gathering evidence: Use of visual security cues in web browsers.
\newblock In {\em Proceedings of Graphics Interface 2005}, pages 137--144.
  Citeseer, 2005.

\end{thebibliography}
\interlinepenalty=10000

\end{document}